\documentclass{article}

\usepackage[utf8]{inputenc}
\usepackage[english]{babel}
\usepackage[backend=biber,style=ieee,citestyle=numeric]{biblatex}
\newrobustcmd*{\citefullauthor}{\AtNextCite{\DeclareNameAlias{labelname}{given-family}}\citeauthor}
\usepackage{authblk}
\usepackage[toc,acronym,translate=babel,acronymlists={hidden},automake]{glossaries-extra}
\GlsXtrEnablePreLocationTag{~-~page:~}{~-~pages:~}
\glssetcategoryattribute{general}{textformat}{emph}
\glssetcategoryattribute{general}{glossnamefont}{emph}
\glssetcategoryattribute{acronym}{glossnamefont}{textbf}
\newglossarystyle{glsstylelong}{
    \setglossarystyle{long}
    
}
\setabbreviationstyle[acronym]{short-long}
\setabbreviationstyle[main]{short-long}
\newglossary[glhidden]{hidden}{glhidden}{glhiddenin}{Hidden Glossary}
\makeglossaries
\loadglsentries{anc/glossary.tex}
\usepackage{csquotes}
\usepackage{xurl}
\usepackage{hyperref}
\usepackage[noabbrev,capitalise]{cleveref}
\usepackage{booktabs}
\usepackage{array}
\usepackage{multirow}

\title{Automated Test Production \\ Complement to ``Ad-hoc'' Testing}
\author[1]{Gomes, J.M.}
\author[1]{Dias, L.A.V.}
\affil[1]{Instituto Tecnológico de Aeronáutica - ITA}

\begin{document}

\maketitle

\begin{abstract}
    A view on software testing, taken in a broad sense and considered a important activity is presented. We discuss the methods and techniques for applying tests and the reasons we recognize make it difficult for industry to adopt the advances observed in academia. We discuss some advances in the area and briefly point out the approach we intend to follow in the search for a solution.
\end{abstract}

\section{Motivation}\label{RS0001:motivacao}

Accepting that tests are important, but are not always implemented or kept up to date during the lifetime of a program, we conclude that nothing has changed since the introduction of the Agile Manifesto earlier this century \cite{beck2001manifesto} which we reproduce below and from which we highlight the passage ``Software that works rather than complete documentation''\cite{beck2001manifesto}.

\begin{displayquote}
    \begin{itemize}
        \item ``\textit{Individuals and interactions over processes and tools}''
        \item ``\textit{Working software over comprehensive documentation}''
        \item ``\textit{Customer collaboration over contract negotiation}''
        \item ``\textit{Responding to change over following a plan}''
    \end{itemize}
\end{displayquote}

This view has come to become an important industry trend \cite{ramesh2006can}\footnote{The 14$^o$ annual report \textit{STATEofAGILE} from \citeyear{one202014th} points out that 95\% of organizations practice agile software development methods. \cite{one202014th}.}, where face-to-face interactions are preferable to formal communication processes and working programs are preferable to comprehensive documentation, leaving the interpretation of the term ``comprehensive'' to each agile development team to decide \cite{hoda2010much}. In fact the agile method suggests that all documentation can be replaced by informal communication with an emphasis on tacit rather than explicit knowledge \cite{cockburn2001agile}.

On the other hand, the adoption of continuous integration and continuous delivery processes and tools has been steadily and unequivocally growing in both industry \cite{feitelson2013development,claps2015journey} and open source projects \cite{hilton2016usage}, which can to some extent be interpreted as a denial of one of the principles of the Agile Manifesto: ``\textit{Individuals and interactions over processes and tools}'', yet this does not come as a relief to the fact that many see benefits in building and maintaining formal models, but are not content to build them as they believe they consume too much time and resources, even believing in the slim chances of success of projects that do not use some modeling \cite{canat2018enterprise}.

The implications of this view for the construction and maintenance of programs and the use and application of development methods and tools are discussed.

\section{Test Production Methods}

The present discussion is a contribution to the understanding of how software testing fits into the present realities perceived by both industry and academia, even if these realities, as we shall see, do not correspond and will not converge. The \gls{TDD} technique is widely cited and recommended by the signers of the Agile Manifesto \cite{beck2001aim}, even though it is not part of the manifesto or its twelve principles \cite{beck2001manifesto}, so we can conclude that the \gls{IT} industry at least recognizes the importance of testing programs. The academia, on the other hand, perceives program testing based on formal specifications as inevitable in pioneering studies since the 1970s \cite{goodenough1975toward,chow1978testing}, the foundations for combining formal methods and program testing being established and accepted, and it is up to the community to put them into practice, optimize and extend them.

In general, we classify the tests in \textbf{Formal}: verifiable by theoretical means or pure logic; and \textbf {Empirical}: verifiable through observation or direct experience\footnote{We take into account the formality of the test and not the conduct of the test, as it is perfectly possible to conduct empirical tests by adopting formal practices in their execution.}.

\subsection{Formal Testing}

\citeauthor{hoare1969axiomatic} and \citeauthor{floyd1986toward} introduced formal methods by introducing the ``Hoare calculus'' for proving the correctness of a program as well as the notions of pre and postconditions, invariants and assertions. His ideas were gradually developed into the current formal software engineering tools and techniques, such as the \gls{OCL} \cite{jos1999object} used to specify constraints in \gls{UML} diagrams.

According to \citeauthor{gaudel2017formal}, for each and every specification method, there is a notation \cite{gaudel2017formal}. Depending on the method, specifications can include expressions in various logical forms, used to write pre and postconditions, axioms of data types, constraints, temporal properties. They can represent definitions of process states, such as:

\begin{itemize}
    \item \gls{CSP} \cite{roscoe1998theory}
    \item \gls{CCS} \cite{milner1984lectures}
    \item \gls{LOTOS} \cite{brinksma1983algebraic}
    \item \gls{Circus} \cite{de2006definiccao}
\end{itemize}
    
Or they can have annotated diagrams, such as:

\begin{itemize}
    \item \gls{FSM} \cite{minsky1967computation}
    \item \gls{LTS} \cite{clarke1986automatic}
    \item Petri Networks \cite{petri2008petri}
    \item etc.
\end{itemize}

But there is more than a syntax. First, there is a formal semantics, in terms of mathematical notions such as:

\begin{itemize}
    \item Predicate transformers for pre and post conditions
    \item Classified sets and algebras for axiomatic definitions
    \item Various types of automata, traces, faults, divergences, for process algebras
\end{itemize}

Second, there is a formal deduction system, making it possible to perform proofs, or other checks (such as model checking), or both. Thus, formal specifications can be analyzed to guide the identification of appropriate test cases.

In addition to syntax, semantics, and the deduction system, formal methods come with some relations between specifications that formalize equivalence or correct step-by-step development. Depending on the context, such relations are called: refinement, conformance, or, in the case of formulas, satisfiability, and are fundamental to test methods \cite{gaudel2017formal}.

\citeauthor{gaudel2017formal} concludes that model-based tests are tests of the \textit{black-box}\footnote{Method of validating functional and external aspects of a computer application.} type, where the internal organization of the program under test is ignored and the strategy is based on a description of the desired properties and behavior of the program\footnote{We separate these tests into a category - that of \textbf{Formal Tests} - that is, tests with a formal basis and that originate from models.}, which may be formal or not, or in other words, these methods target certain classes of faults and assume that the program is exempt from other types and classes of faults \cite{gaudel2017formal}.

\subsection{Empirical Tests}

Without formal defined specifications \textit{a priori}, which as we have seen in ``\nameref*{RS0001:motivacao}'' is a trend in the industry, we are left only with informal and empirical practice\footnote{Which generally means: verifiable by direct observation or experience rather than by theory or pure logic, even though it is possible to adopt formal practices during an empirical procedure.} for the verification and validation of the correctness of computer program implementation\footnote{Nothing prevents that, even starting from a basis of formal specifications, empirical tests be adopted in the verification of the implementation.}. One of the practices advocated by supporters of agile methods is \gls{TDD}, where tests are written even before the program itself, but it does not show clear benefits\footnote{The practice of \gls{TDD} is advocated mainly because the alternative is to have no tests at all after the program is ready\cite{george2004structured}.} compared to the option of implementing the tests after the program is ready \cite{shull2010we, josefsson2004making,madeyski2010impact}, or it may be linked to the fact that processes like \gls{TDD} encourage stable and refined steps of continuous improvement \cite{fucci2016dissection}.

In the informal test, we have a relation of the hypothesis to an observation statement, which is nothing more than a proposition about the perceptible properties of some entity, set of entities, or system, followed by a rule transmission where, if the observation statement directly confirms the hypothesis, then indirectly it confirms any of its logical consequences \cite{hempel1945studies}.

We can state that formal tests are cases of inductive inference\footnote{We cannot call ``formal tests'' a case of ``indirect confirmation''.}, and that in empirical tests we have a direct confirmation of the hypothesis, but without the soundness and precision that formal methods\footnote{Formal methods pursue qualitative and quantitative metrics of the soundness and precision of the method itself.} guarantee \cite{iso1991information} as a consequence of the \textit{ad hoc} attitude with which the informality of design\footnote{And as we said earlier, not necessarily of the actual conduct, which can be perfectly formal.} of empirical testing is practiced.

Just as using only \textbf{Formal Methods} we are unable to judge all the possibilities of flaws that a program may present\cite{dijkstra1979structured}, we can state that \textbf{Empirical Methods} are also so, and for the same reasons, with the aggravating factor of introducing a certain randomness\footnote{The observer's objectivity and his judgment.} to the process.

\subsection{Static and Dynamic Analysis}

This is a case where the test can either be defined \textit{a priori} (as in \gls{TDD} or model-based) or \textit{a posteriori} (as most informal tests are done), and which according to \citeauthor{gaudel2017formal}, would be the answer to the lack of coverage of Formal Tests, but which as we will see below, also present problems of application in practice.

Static analysis was introduced in \citeyear{osti_5361454} with the work \citetitle{osti_5361454} by \citeauthor{osti_5361454} \cite{osti_5361454}. The nature of verification performed by static parsers include \cite{aiken1995static,ayewah2008using} (but not limited to only these) the following analyses:

\begin{itemize}
    \item \textit{Layout} and source code formatting
    \item Identifying language constructs known to be non-portable
    \item Identifying algorithm constructs known to be unsafe
    \item Use of variables or constants with suspect names and contents (for example: \textbf{PASSWORD = 'SECRET'})
    \item Detection of faults not considered by compilers
    \item Control flow analysis (detection of \textit{loops})
    \item Detect data usage in variables before a value has been entered
    \item Detect value overloading in variables (assign a very large value to a variable that only supports small values - in some languages assign a \textbf{DOUBLE} value to a simple \textbf{INT} variable)
    \item Detect memory overflow (leak) or the non-validation of may memory overflow (assigning a very long constant to a variable that supports a small memory size)
    \item Detect leakage of \textit{handles} (the reference to the control structure) of files and accesses to communication resources
    \item Check permission to perform certain operations
    \item Ensuring the termination of a processing (or ensuring indications that it will not terminate)
    \item Ensure the order in which processing is performed and terminated in a way that maintains the integrity of the information (or ensure that it gives indications that the information is not intact)
    \item Ensure that the process can be observed as deterministic\footnote{If an action is visible to the environment (i.e. if it performs data retrieval or changes data), then we say it is observable. The order of execution of non-priority rules will make a difference in the order of appearance of observable actions.} (or ensure that there are indications that the process cannot be observed as deterministic)
\end{itemize}

Many of these validations can be (and most often are) done by compilers (when the language is compiled)\cite{wilson1995efficient}. Since the purpose of the compiler is to generate executable code and not to check for programming faults, and other classes of faults can only be determined at runtime, such as memory overflow, which only occurs if a very long constant is supplied during program use,\footnote{Although it is possible, as we can see later, to predict overflow using one of the many static analysis methods available.}, then specialized checkers such as \textbf{Linters}\cite{darwin1988checking} are adopted. Capable of detecting a wide range of faults, including style (\textit{layout} of source code), some use source code annotations to achieve better problem detection, at the expense of extra developer work \cite{evans1996static,jackson1995aspect,detlefs1996overview,detlefs1998extended,jensen1997automatic}.

Static analyzers can then be classified (see \cref{RS0001:tab:classif_estatic}) into various types and capabilities, covering the detection of several possible fault categories, from implementation to vulnerability and security related.

\begin{table*}\scriptsize
    \centering
    \begin{tabular}{ >{\centering\arraybackslash}p{20mm} p{110mm} }
        \toprule
            \textbf{Classe} & \textbf{Descrição} \\
        \midrule
            \tiny{\textbf{Lexical Analysis}} &\footnotesize{\textit{Lexical analysis is based on the grammatical structure of the language. It divides the program into small parts that are compared to known fault libraries. Disregarding syntax, semantics and interaction between subroutines, the incidence of false positives is high \cite{li2010comparative}.}} \\
        \hline
            \tiny{\textbf{Type Inference}} &\footnotesize{\textit{It infers the type of variables and functions by the compiler or interpreter, and checks that accesses to these variables and functions conform to predefined rules for the type \cite{hankin1994deriving}.}} \\
        \hline
            \tiny{\textbf{Data Flow Analysis}} &\footnotesize{\textit{Refers to collecting semantic information from source code, and using algebraic method determines the definition and use of variables at compile time. Starting from the execution flow graph, a data flow analysis determines whether values in a program are flagged as potentially vulnerable variables \cite{fosdick1976data}.}} \\
        \hline
            \tiny{\textbf{Rule Checking}} &\footnotesize{\textit{Checks the security of a program using pre-set rules \cite{hayes1985rule}. Some rules, such as requiring execution under elevated privilege, carry security implications \cite{li2010comparative} and are detected.}} \\
        \hline
            \tiny{\textbf{Constraints Analysis}} &\footnotesize{\textit{Divided between constraint generation and constraint resolution during the analysis process. Constraint generation sets variable types or analyzes the constraint system between different states of execution using predetermined rules; constraint resolution applies and resolves the generated constraints \cite{li2010comparative}.}} \\
        \hline
            \tiny{\textbf{Comparison of Correction Snippets}} &\footnotesize{\textit{Comparison of source or binary code snippets changed during the process of fixing flaws is used to find known implementation gaps. After patches have been applied to a program, the comparison serves to determine the location and causes of the vulnerability to which they apply \cite{li2010comparative}.}} \\
        \hline
            \tiny{\textbf{Symbolic Execution}} &\footnotesize{\textit{It represents program inputs as symbols instead of the actual data, and produces algebraic expressions over the symbol in the implementation process. By the constraint solving method symbolic execution can detect possible failures \cite{boyer1975select,king1976symbolic,howden1976experiments,clarke1976program}.}} \\
        \hline
            \tiny{\textbf{Abstract Interpretation}} &\footnotesize{\textit{It is a formal description of program analysis, which maps the program to abstract domains. The technique requires completeness, which makes it impractical for very large programs, but proves correct for all possible inputs \cite{abramsky1987abstract,nielson1994abstract}.}} \\
        \hline
            \tiny{\textbf{Proof of Theorems}} &\footnotesize{\textit{Semantic analysis of the program, which can solve infinite state system problems \cite{davis2001early,bibel2007early}. First convert the program into a logical formula, then prove that the program is a valid theorem using axioms and rules \cite{li2010comparative}.}} \\
        \hline
            \tiny{\textbf{Model Verification}} &\footnotesize{\textit{Starting from formal models, such as state machines or directed graphs, it runs through them and compares the model with the implementation to see if it matches the characteristics predefined by the first \cite{clarke2018model}.}} \\
        \bottomrule
    \end{tabular}
    \caption{Classification of Static Analyzers}
    \label{RS0001:tab:classif_estatic}
\normalsize\end{table*}

The problem with static analyzers is the high false positive rate (alerts that are not real problems), low understandability of alerts and lack of automation in quick fixes for the large number of identified problems \cite{johnson2013don}, such as: \cite{panichella2015would} code structure and \cite{zampetti2017open} coding patterns, which could easily be fixed using automatic refactoring techniques \cite{agnihotri2020systematic}, but as we will see below, the available tools are not in line with the latest advances made by the scientific community.

Dynamic analysis, on the other hand, is in contrast to static analysis and contemplates the forms best known and adopted by the industry in the application of software testing \cite{myers2004art} (see \cref{RS0001:tab:classif_dynamic}).

\begin{table*}\scriptsize
    \centering
    \begin{tabular}{ >{\centering\arraybackslash}p{20mm} p{110mm} }
        \toprule
            \textbf{Classe} & \textbf{Descrição} \\
        \midrule
            \tiny{\textbf{Unit Test}} &\footnotesize{\textit{The process of testing subprograms, subroutines, classes, or functional units within a program to verify that there are no programming flaws \cite[p.~486]{iso2017iso}.}} \\
        \hline
            \tiny{\textbf{Integration Test}} &\footnotesize{\textit{Testing phase where the functional units are combined and tested as a group to assess whether they worked properly in the complete system \cite[p.~235]{iso2017iso}.}} \\
        \hline
            \tiny{\textbf{System Testing}} &\footnotesize{\textit{Test conducted on multiple integrated systems to evaluate their ability to communicate with each other and achieve general and specific integration requirements \cite[p.~545]{iso2017iso}.}} \\
        \hline
            \tiny{\textbf{Acceptance Test}} &\footnotesize{\textit{Testing of a system or functional unit generally performed by the buyer or user on-site after installation of the software to make sure that the contractual requirements have been met \cite[p.~5]{iso2017iso}.}} \\
        \bottomrule
    \end{tabular}
    \caption{Classification of Dynamic Tests}
    \label{RS0001:tab:classif_dynamic}
\normalsize\end{table*}

\section{The Challenge of Testing}

\subsection{Software Quality}

\citefullauthor{hoare1996did} in the research \citetitle{hoare1996did} conducted in \citeyear{hoare1996did} states that it was reasonable to predict that the size and ambition of software products would be severely limited by the lack of reliability in their components. Estimates suggested, in its study, that professionally written programs may contain between one and ten correctable faults for every thousand lines of code; and any one software fault, in principle, can have a spectacular effect (or worse: a subtly misleading effect) on the behavior of the entire system \cite{hoare1996did}.

\citeauthor{hoare1996did} found at the time that the software patch problem turned out to be far less serious than anticipated. An analysis by Mackenzie \cite{mackenzie1994computer} showed that of several thousand deaths attributed to computer applications, only ten or so could be explained by software crashes: most due to a few cases of incorrect dosage calculations in radiation cancer treatment. Similarly, predictions of collapse due to the size of computer programs have been falsified by the continuous operation of real-time software systems now measured in tens of millions of lines of code and subject to thousands of updates per year.

In his review \citeauthor{hoare1996did} concludes that, despite appearances, modern software engineering practice owes much to the theoretical concepts and ideals of early research in this field; and that formalization and proof techniques have played an essential role in the validation and progress of research.

\citeauthor{hoare1996did} concludes that the main factors for the apparent success of the software are:

\begin{itemize}
    \item \textbf{Management} - \textit{The most dramatic advances in the delivery of reliable software are directly attributable to a wider recognition of the fact that the process of program development can be predicted, planned, managed, and controlled just as in any other branch of engineering.}
    \item \textbf{Test} - \textit{Thorough testing is the cornerstone of reliability in quality assurance and control in modern production engineering. Tests are applied as early as possible throughout the production line. They are rigorously designed to maximize the probability of detecting failures and as quickly as possible.}
    \item \textbf{Debugging} - \textit{The secret of successful testing is that it checks the quality of the process and methods by which the code was produced. But there is an entirely different and very common response to the discovery of a flaw by testing: simply fix it and get on with the job. This is known as debugging, by analogy with trying to get rid of a mosquito infestation by killing the ones that bite - much faster, cheaper and more satisfying than draining the swamps in which they breed.}
    \item \textbf{Excess Engineering} - \textit{The concept of safety factor is very widespread in engineering. After calculating the worst case load on a beam, the civil engineer will try to build it at least twice as strong. In computing, a continuous drop in the price of storage and increased processing power has made it acceptable to add redundancies to reduce the risk of software failures and a smaller scale of damage. This leads to the same kind of over-engineering required by law for bridge construction; and it is extremely effective, although there is no clear way to measure it by a numerical factor.}
    \item \textbf{Programming Methodologies} - \textit{Most of the measures described so far for achieving reliability in software are the same ones that have been proven equally effective in all engineering disciplines. But the best general techniques for management, quality control, and safety would be totally useless by themselves; they are only effective when there is a general understanding, a common conceptual framework and terminology for discussing the relationship between cause and effect, between action and consequence. Research in programming methodology has this goal: to establish a conceptual framework and a theoretical basis to assist in the systematic derivation and justification of each design decision by a rational and explicable line of reasoning.}
\end{itemize}

\subsection{Perceived Quality when Using Software}

According to the \gls{NIST} report, the estimated impact (in the United States) of inadequate software testing infrastructure is $859$ billions dollars and the potential cost savings from feasible improvements is $822$ billions dollars. Software users account for a larger share of the total costs of inadequate infrastructure (64 percent) compared to ``viable'' cost reductions (52 percent) because a large share of user costs are due to prevention activities. Whereas mitigation activities decrease proportionally to the decrease in the number of failures, prevention costs (such as redundant systems and investigating purchasing decisions) are likely to persist, even if only a few errors are expected. For software developers, the feasible cost savings are approximately 50 percent of the total costs of inadequate infrastructure. This reflects a more proportional decrease in testing effort as testing resources and tools improve \cite{planning2002economic}.

\begin{table*}\scriptsize
    \centering
    \begin{tabular}{ p{30mm} >{\bfseries\centering\arraybackslash}p{20mm} >{\bfseries\centering\arraybackslash}p{15mm} >{\bfseries\centering\arraybackslash}p{20mm} >{\bfseries\centering\arraybackslash}p{20mm} >{\bfseries\centering\arraybackslash}p{20mm}  }
        \specialrule{.1em}{.05em}{.05em} 
            &
            \begin{tabular}[b]{>{\bfseries\centering\arraybackslash}p{20mm} }
                Testers / Employees (millions)
            \end{tabular}
            & 
            \begin{tabular}[b]{>{\bfseries\centering\arraybackslash}p{35mm} }
                Cost of inadequate testing infrastructure
            \end{tabular}
            &  &                   
            \begin{tabular}[b]{>{\bfseries\centering\arraybackslash}p{40mm} }
                Potential cost reduction with feasible improvements
            \end{tabular}
            &  \\ 
        \cline{3-6}
            &  & Unit cost & Total cost (million US\$) & Unit cost (million US\$) & Total Cost (million US\$) \\ 
        \hline
            Developers & 0,302 & 69.945 & 21.155 & 34.964 & 10.575 \\                 
            Users &  &  &  &  &  \\ 
            \setlength\parindent{24pt}\par{Industry} & 25,0 & 459 & 11.463 & 135 & 3.375 \\ 
            \setlength\parindent{24pt}\par{Services} & 74,1 & 362 & 26.858 & 112 & 8.299 \\ 
        \hline
            Total &  &  & 59.477 &  & 22.249 \\ 
        \specialrule{.1em}{.05em}{.05em} 
    \end{tabular}
    \caption{Estimated national impact in the US (adapted from \cite{planning2002economic})}
    \label{RS0001:tab:nist}
\normalsize\end{table*}

If we add up everything from minor inconveniences in our daily lives to incalculable human and social damage from software failures, the perception we have may be quite different from that of \citeauthor{hoare1996did} in his study. This is because today the penetration of computerized systems in our lives, with its own challenges and opportunities due to the great convergence of connected systems, interoperability and massive distribution of information, can make the most insignificant failure from a mere annoyance (such as losing access to your favorite music playlist) to a catastrophe of global proportions (such as a widespread failure in a worldwide satellite communications system).

\subsection{The Gap Between Industry and Scientific Advances}

In \citeyear{hoare1996did} \citeauthor{hoare1996did} noted that academic research gains in programming methodologies took up to 20 years to be adopted by industry as a sign of maturity and sanity - only in very specific areas and for a brief period would it be justified to apply the latest pure research advances to people's everyday lives \cite{hoare1996did}. This mismatch also has the benefit of providing adequate planning of research and education as well as adequacy of the installed park in the industry. The result of not following this step is to adopt immature technologies and practices, with unpredictable and undesirable results, with no skilled labor available to apply it and make the necessary corrections when failures occur\footnote{If this scenario sounds like something that is happening in your industry, then maybe this is the reason.}.

Another consequence of not observing the maturity of cutting-edge research before its adoption in practice is the fact that, paradoxically, mature and effective technologies have not yet been adopted by industry, or when they are, they are isolated cases that cause astonishment when they present better results than those obtained with ``state-of-the-art technologies''. As an example we cite the adoption of the \textit{pairwise} technique for test generation. The mathematical theory behind this technique has been around since the 1960s (see \citetitle{webb1965design} publishied in \citeyear{webb1965design} \cite{webb1965design}), the application in software testing using \textit{pairwise} was presented earlier this century (see \citetitle{du2000combinatorial} published in \citeyear{du2000combinatorial} \cite{du2000combinatorial}). Recent research using these techniques (see \cref{RS0001:tab:research_pairwise}) shows promising numbers\footnote{We are aware that this sampling is neither meaningful nor representative, but only illustrative from our point of view.}:

\begin{table*}
    \centering
    \begin{tabular}{ >{\raggedright\arraybackslash}p{70mm} p{10mm} >{\raggedleft\arraybackslash}p{10mm} >{\raggedleft\arraybackslash}p{10mm} >{\raggedleft\arraybackslash}p{10mm} }
        \toprule
            \textbf{Research} & \textbf{Type} & \textbf{Tests} & \textbf{Time} & \textbf{Defects} \\
        \midrule
            \multirow{2}{*}{\parbox{70mm}{\citetitle{monteiro2014case} \cite{monteiro2014case}}} & \textit{Ad-hoc} & 14,041 & 20h & 10 \\
                                                                                    \cline{2-5} & \textit{Pairwise} & 68 & 4h & 10 \\
        \hline            
            \multirow{2}{*}{\parbox{70mm}{\citetitle{da2015case} \cite{da2015case}}} & \textit{Manual} & 159 & 6h & 3 \\
                                                                        \cline{2-5} & \textit{Pairwise} & 17 & 1h & 3 \\
        \bottomrule
    \end{tabular}
    \caption{Pairwise Application Research Results}
    \label{RS0001:tab:research_pairwise}
\end{table*}

With results like this, it was expected that the adoption of the \textit{Pairwise} technique to tests production in a cost-effective way would be more welcomed by the industry\footnote{Informally, in our contacts with software development practitioners and testing experts and discussions about the practice of \textit{Pairwise} have ranged from ignorance of its existence to negative concepts and objections to its use as ineffective.}.

\section{Promises of Formal Development}

\subsection{Model Driven Development}

One of the most promising approaches to computer program development was \gls{MDD} and \gls{MDA}, where models are the primary artifacts and the others, such as code, are generated from them \cite{yusuf2006implement}. The goal is to raise the level of abstraction, making software development closer to solving the requirements and problems outlined by its future users and making the developer's life simpler and easier \cite{hailpern2006model} and providing mainly automation of the process \cite{jacobs_arcast_nodate}. According to \citeauthor{yusuf2006implement} and \citeauthor{swithinbank2005patterns}, the advantages of using \gls{MDD} are:

\begin{itemize}
    \item Increased developer productivity - because of automation and focus on requirements analysis
    \item Ease of maintenance - many software was developed by specialists who left the organization at some point, and the technique would facilitate the evolution by retaining the knowledge of these specialists
    \item Legacy reuse - can make it easy and feasible to migrate old applications to new systems by applying the technique
    \item Adaptability - adding or modifying is made easy given the automation already in place
    \item Consistency - every application will strictly follow the pattern established by the tools
    \item Repetition - great return on investment if applied throughout an organization
    \item Improved communication with sponsors - models are easier to interpret than code
    \item Improved project communication - templates help to understand the system design and assist in the discussion about the system itself
    \item Domain knowledge capture - if there is sufficient documentation of the system, the organization's knowledge is maintained
    \item Long-term asset - high-level models and abstractions of business solutions are immune to technological change
    \item Ability to postpone technology decisions - focus on solving business problems allows decisions on non-functional problems to be left for a more opportune time
\end{itemize}

\subsubsection{Problems with models}

The biggest problem with using models as the only source for software production is that trying to solve an organizational problem from conceptual abstractions larger than the machine languages used by computers to run programs implies a reduction of information \cite[p.~90]{langer1953feeling}. This information has to be supplanted by the \gls{MDD} tool itself by means of ready-made patterns, or from the developer by means of extensions, and that leads, according to \citeauthor{hailpern2006model} \cite{hailpern2006model} to other problems:

\begin{itemize}
    \item Redundancy - because of the widespread use of ready-made code examples
    \item Unbridled back and forth problems - to adjust the model to conform to another system or module
    \item Moving complexity elsewhere rather than reducing it, requiring even more specialization
\end{itemize}

\subsubsection{Future of MDD}

Standardization around \gls{UML} and tool interoperability around the \gls{XMI}\cite{omg_xml_2015} standard can lead the open source community to produce products that can leverage development using \gls{MDD}. Tools such as the Eclipse Modeling Framework (see \url{https://www.eclipse.org/modeling/emf/}) is an example of technology with this kind of potential, however this leads us to another conclusion.

\subsubsection{Prospects}

Our view is that, the main barrier to the adoption of technologies like \gls{MDD}, is how quickly this kind of solution becomes irrelevant.

This irrelevance happens as the application and use of information technologies and platforms evolve.

In the 1970s and 1980s, the adoption of \gls{CASE} tools, which we can say were the precursors of \gls{MDD} and \gls{MDA}, was seen as a solution to the same problems we have listed above. At that time software development took place mainly on large computers, the \textit{Mainframes}. But at the same time personal computers emerged, which at first were not seen as business tools, this soon became an untruth with the release of the IBM PC in 1981\cite{miller_why_2011} and since then software development has moved from the older and more expensive platform (\textit{Mainframes}) to the more modern and cheaper (\glspl{PC}), and this became increasingly true with the adoption of local networks like Novell in 1979 \cite{proven_how_2013} with over $500.000$ computers installed in the world \cite{payne2006ccie} at the time. This movement continued, but once again changed focus. In 1989 Tim Berners-Lee invented the World Wide Web, in 1993 we had the release of the Mosaic browser by \gls{NCSA}, and in 1994 we had Netscape Navigator created by the same developers, now in a private company of the same name. Since then the development has been turning to applications presented by the browsers but running on corporate servers on the Internet. In early 2007 Apple introduces the iPhone, and at the end of the following year Google introduces Android. Still supported by the basic Internet infrastructure, application development shifts focus once again to the new mobile platform. And these days, some technologies are on the threshold, or at least promise to be, of creating new platforms, and among them we can mention Bitcoin (announced in 2009), virtual reality (as used in airplane pilot training and introduced as a consumer product in the 1990s by computer game companies like Sega in 1991) and augmented reality (made popular in games like Pokémon Go in 2016) and finally the renaissance of Artificial Intelligence with the adoption of Machine Learning techniques.

This rapid evolution and shift of focus to different platforms, with different approaches that decisively impact the architecture of the systems, databases, operating systems, programming languages, forms of presentation, number of application layers, and different \glspl{API} employed to mediate an increasingly large and complex network of interconnected products and services makes it practically impossible to develop, train personnel, and make them productive in the employment of any technology with the nature of the \glspl{MDD} tools, which end up being relegated only to the role of modeling, right at the initial requirements gathering phase, within a longer development life cycle and without fulfilling the promise of covering it completely that has been made since the 1970s and 1980s by the \gls{CASE}\cite{mercurio1990ad} tools, and which, as we saw earlier in this introduction, often does not motivate software development professionals and decision makers to bear the cost and time required in their absortion and deployment.

\section{Conclusions and Future Work}\label{RS0001:conclusao}

If in one hand we have the promise of great advances and improvements in the quality of software products by applying techniques and tools developed by both academia and industry, despite the expected (and even desirable) delay between the development and adoption of these new technologies, we also have on the other hand the adoption of practices by the industry that make it difficult to incorporate certain mature technologies, or even to put them to the test, due to the lack of formalization that these practices prescribe in the name of agility in producing products quickly and meeting the desires of their customers.

Without the adoption of formal software development methods, it is not possible to continue and progress with the advanced quality methods and methodologies developed in academia.

The solution to this would be a back-and-forth approach, whereby by reverse engineering and starting from the source code of the computer programs, formal models are deduced and then complemented by the developers in order to produce the artifacts and inputs necessary for formal methods of quality verification and validation. Automation and adoption of standards are key to keeping costs within acceptable parameters for the industry.

This approach has its pros and cons. Using reverse engineering to produce formal models will cause loss of information \footnote{In general, models should have less information than the finished products that originated from them}, and this and other problems to come are what we set out to address.

We intend to continue these studies with an analysis of the State of the Art in the conception and production of computer program tests, followed by ways of bringing together the methods and practices adopted by industry and the techniques developed by academia.

\clearpage
% \printglossary[type=main,style=glsstylelong]
% \bigskip
\printglossary[type=acronym,style=glsstylelong]

\printbibliography[heading=bibintoc]

\end{document}